\documentclass[12pt,preprint]{aastex}

\input psfig.sty

\slugcomment{\bf Accepted for Publication in ApJ}

\shorttitle{Lensing by Proxima}
\shortauthors{Sahu et al.}

\begin{document}

\def\degpoint{^\circ\mskip-7.0mu.\,}
\def\feh{{\rm [Fe/H]}}
\def\gtabout{\, {}^>_\sim \,}
\def\Hipp{{\it Hipparcos}}
\def\HST{{\it HST}}
\def\IUE{{\it IUE}}        
\def\kms{{\>\rm km\>s^{-1}}}
\def\ltabout{\, {}^<_\sim \,}
\def\secpoint{''\mskip-7.6mu.\,}
\def\smallstrut{\hbox{\vrule height 0.3em depth 0em width 0pt}}
\def\Teff{T_{\rm eff}} 
%



\title{Microlensing Events  by Proxima Centauri in 2014 and 2016:  
Opportunities for Mass
Determination and Possible Planet Detection\altaffilmark{1}}




\author{Kailash C. Sahu\altaffilmark{2},
Howard E. Bond\altaffilmark{2,3}, Jay Anderson\altaffilmark{2}, and 
Martin Dominik\altaffilmark{4}
}



\altaffiltext{1}
{Based in part on observations made with the NASA/ESA {\it Hubble Space
Telescope}, obtained at the Space Telescope Science Institute, which is operated
by the Association of Universities for Research in Astronomy, Inc., under NASA
contract NAS 5-26555.} 

\altaffiltext{2}{Space Telescope Science Institute, 3700 San Martin Drive,
Baltimore, MD 21218, USA; ksahu@stsci.edu, jayander@stsci.edu }

\altaffiltext{3}{Department of Astronomy \& Astrophysics, Pennsylvania State
University, University Park, PA 16802, USA; heb11@psu.edu}

\altaffiltext{4}{Royal Society University Research Fellow,
SUPA, University of St.\ Andrews, School of Physics \&
Astronomy,  North Haugh, St.\ Andrews, KY16 9SS, UK;
md35@st-andrews.ac.uk
}


\begin{abstract}

We have found that Proxima Centauri, the star closest to our Sun, will pass
close to a pair of faint background stars in the next few years.  Using {\it
Hubble Space Telescope\/} (\HST\/) images obtained in 2012 October, we determine
that the passage close to a mag 20 star will occur in 2014 October (impact
parameter $1\secpoint 6$), and to a mag 19.5 star in 2016 February (impact
parameter $0\secpoint 5$). As Proxima passes in front of these stars, the
relativistic deflection of light will cause shifts in the
positions of the background stars by $\sim$0.5 and 1.5~mas, respectively,
readily detectable by \HST\/ imaging, and possibly by {\it Gaia\/} and
ground-based facilities such as VLT\null. Measurement of these astrometric
shifts offers a unique and direct method to measure the mass of Proxima.
Moreover, if Proxima has a planetary system, the planets may be detectable
through their additional microlensing signals, although the probability of such
detections is small. With astrometric accuracies of 0.03~mas (achievable with
\HST\/ spatial scanning), centroid shifts caused by Jovian planets are
detectable at separations of up to $2\secpoint0$ (corresponding to 2.6~AU at the
distance of Proxima), and centroid shifts by Earth-mass planets are detectable
within a small band of  8 mas (corresponding to 0.01~AU) around the source
trajectories.  Jovian planets within a band of about 28 mas  (corresponding to
0.036 AU) around the source trajectories would produce a brightening of the
source by $>$0.01~mag and could hence be detectable. Estimated timescales of the
astrometric and photometric microlensing events due to a planet range from a few
hours to a few days, and both methods would provide direct measurements of the
planetary mass.

\end{abstract}


\keywords{Gravitational lensing: micro, Astrometry, Stars: individual:
Proxima Centauri, planetary systems}



\section{\bf Proxima Centauri: Our Nearest Neighbor}

Proxima Centauri ($\alpha$~Centauri~C, V645~Cen, GJ~551; \Hipp\/ parallax
$771.64\pm2.60$~mas) is the nearest known star, lying 1.30~pc from the Sun. It
is located at a projected angular separation of $2\fdg2$ and a linear distance
of 0.07~pc (14,500~AU) from the bright and slightly more distant binary
$\alpha$~Cen A+B\null. The distance, proper motion, and radial velocity of
Proxima are closely similar to those of $\alpha$~Cen A+B, showing that the
system forms a wide triple (Wertheimer et al.\ 2006); however, Proxima's large
separation from $\alpha$~Cen A+B means that it has evolved effectively as a
single star. Proxima is a low-mass, M5~Ve dwarf and flare star, with an apparent
magnitude of $V = 11.05$, a radial velocity of $-22.4\,\rm km\, s^{-1}$ (Torres
et al.\ 2006), and a proper motion of  
$3\secpoint8526 \pm 0\secpoint0026 \,\rm yr^{-1}$ in position
angle (PA) $281\fdg461 \pm 0\fdg036$,  as measured by \Hipp\/ (van~Leeuwen 2007; also see
Benedict et al.\ 1999\footnote{Benedict et al.\ 1999 give a proper-motion PA of
$78\fdg46$, evidently measuring it from north through west instead of the
conventional direction.})

\subsection{Mass of Proxima Centauri}

Proxima, as an M dwarf, represents the most common type of star. Proxima's
radius is known from VLTI (Very Large Telescope Interferometry) observations to
be $0.145\, R_\odot \pm0.011 \, R_\odot$ (S\'egransan et al.\ 2003). But, as an effectively isolated
star, Proxima's mass can only be estimated from mass-luminosity (M-L) relations
as $\sim$$0.12\,M_\odot$. 

A star's mass is its most important physical parameter, being the primary
determinant of its temperature, radius, luminosity, lifetime, and ultimate fate.
The only directly measured stellar masses come from double-lined eclipsing
systems and visual binaries. For dM stars like Proxima, the observational
situation is particularly poor, with less than a dozen double-lined eclipsing
systems known that have primary masses of $<$$0.6\,M_\odot$ (e.g., Kraus et al.\
2011). Moreover, the eclipsing systems are not representative of field dM stars,
because synchronous rotation causes stellar activity, starspots, and inflated
radii. The masses of single low-mass stars have to be inferred largely from
theoretical M-L relations, which suffer from poorly understood opacities and
interior structure (e.g., Chabrier \& Baraffe 1997, 2000). 

\subsection{Planets around Proxima Centauri}

Ground-based radial-velocity (RV) measurements have recently led to discovery of
a $m \sin i = 1.1\,M_{\rm Earth}$ planet orbiting $\alpha$~Cen~B at a distance
of 0.04~AU (Dumusque et al.\ 2012). However, all attempts to detect planets
around Proxima Centauri have so far been unsuccessful.  M dwarfs are generally
not optimal for search for planets though RV techniques, since
a large fraction of them have surface activity, which increases RV jitter (e.g.,
Campbell et al.\ 1991; Wright 2005).  However, Proxima itself has a relatively
small RV scatter of 3.1~m\,s$^{-1}$, and RV techniques have been used
extensively to search for planets around it. These data exclude the presence of
any planet in a circular orbit with $m \sin i > 1 \, M_{\rm Neptune}$ at orbital
separations of $a < 1$ AU (Endl \& K\"urster 2008). However, this does not rule
out the presence of larger planets whose orbital plane may be close to face-on. 
No planetary transits have been reported for Proxima.  Astrometric techniques
have also been unsuccessful in detecting any planets around Proxima, which
excludes the presence of any planet with mass $>$$0.8 \,M_{\rm Jup}$ in the
period range 1 day $<P<1000$ days, corresponding to orbital separations of 0.01
to 1 AU (Benedict et al.\ 1999).

\section{\bf Predicting Astrometric Microlensing Events Due to Nearby Stars}

In this paper we discuss an alternative method for measuring masses of nearby
single stars and for searching for planetary companions, and an upcoming
opportunity to apply this method to Proxima Centauri. The special importance of
microlensing events produced by very nearby stars was emphasized by Paczy\'nski
(1996). For such events the term ``mesolensing'' has been suggested (Di~Stefano
2008), because the  large angular Einstein radii of nearby stars, coupled with
their typically large proper motions, yield relatively large probabilities of
such events occurring. Searches for upcoming close stellar passages of
high-proper-motion stars near background sources have been carried out by Salim
\& Gould (2000) and Proft, Demleitner, \& Wambsganss (2011), who have published
lists of candidate events (see also L\'epine and Di~Stefano 2012, who discuss a
specific event). 

In the case of microlensing events due to nearby stars, the foreground star will
almost always be much brighter than the background source, making it difficult to
measure its astrometric shift of the fainter background star.  The contamination
from the foreground star reduces when the lens-source separation is larger, but then
the astrometric signal also reduces, making the measurement again difficult.  
Measuring the astrometric signal will be easiest if we can find a situation where
the centroid shift is  large even when the lens--source angular separation is
large. Such a situation arises for the very nearest stars, for which the angular
Einstein radii are the largest.

To search for such very favorable events, we started with the Luyten
Half Second (LHS) catalog  (Luyten
1979), a compilation of all known stars with proper motions greater than
$0\secpoint5 \,\rm yr^{-1}$.  The coordinates in the LHS catalog are only
approximate, so we used improved positions and proper motions determined by
L\'epine \& Shara (2005) (mostly for northern-hemisphere stars), and the revised
positions and proper motions published by Bakos, Sahu, \& Nemeth (2002) for the
remainder.   Unlike the previous studies, we took the parallaxes of the LHS stars
into account, if they are available in the SIMBAD database, since parallax
significantly alters the circumstances of the events for very nearby stars. We then
projected the positions of all $\sim$5,000 LHS stars forward over the next 40~years
and searched for close passages (impact parameter $<$$2''$) near background stars
contained in the GSC~2.3 catalog\footnote{{\tt
http://gsss.stsci.edu/Catalogs/GSC/GSC2/gsc23\slash gsc23\_release\_notes.htm}~.
The Guide Star Catalogue--II is a joint project of the Space Telescope Science
Institute and the Osservatorio Astronomico di Torino. Participation of the
Osservatorio Astronomico di Torino is supported by the Italian Council for Research
in Astronomy. Additional support is provided by European Southern Observatory,
Space Telescope European Coordinating Facility, the International Gemini project,
and the European Space Agency Astrophysics Division.} (Lasker et al.\ 2008). 

\section{\bf The Proxima Centauri Events of 2014 and 2016}

Of the events that we have found will occur in the next few years, one of the
most interesting is the close passage of Proxima Centauri in front of a pair of
background source stars of $V$ magnitudes 20 and 19.5, separated by $4\secpoint3$. 
Closest approaches will occur in 2014 and 2016, respectively.  Figure~1 shows
images of the Proxima field from the Digitized Sky Survey (DSS)\footnote{The
Digitized Sky Survey was produced at the Space Telescope Science Institute 
under U.S. Government grant NAG W-2166.} IR and  2MASS (Cutri et al.\ 2003)
surveys, taken in 1997 and 2000, respectively. Proxima is the bright star to the
left, marked with a blue circle, and moving rapidly to the right
(west-northwest). A second green circle marks the pair of background stars. 

Although it might appear from a cursory inspection of the ground-based images in
Figure~1 that close passages must occur frequently, passages as close as
$\sim$$1\secpoint5$ are actually rare. Moreover, Proxima is
exceptional in having an Einstein ring radius as large as 28~mas, because it is
so close to the Earth. For more typical proper-motion stars at considerably
larger distances than Proxima, only extremely close passages ($<$$0\secpoint3$)
are of interest from a microlensing standpoint---and they would generally suffer
considerably from large differences in brightnesses.

To refine the impact parameters and times of closest approach, we obtained {\it
Hubble Space Telescope\/} (\HST\/) images of the field on 2012 October~1.
Observations were taken with the UVIS channel of the Wide Field Camera~3 (WFC3)
in the F475W, F555W, F606W, and F814W filters. Since the background stars are
$>$8~mag fainter than Proxima, we obtained two sets of observations, a short 
one with an exposure time of 0.5~s (the minimum allowed exposure time with WFC3)
in each filter, and a long one with an exposure time of 100 to 200~s.  An extra
set of short and long exposures were obtained in the F555W filter, which were
dithered by about $4''$ with respect to the first set. The two long exposures
obtained in the F555W filter were used to produce a cosmic-ray cleaned image of
the field.   Since Proxima is heavily saturated in the longer exposures, care
was taken to choose a telescope orientation such that neither the diffraction
spikes nor the charge bleeding from Proxima would affect the background
sources.   Proxima was not saturated in the short exposures in F475W an F555W
filters, but was slightly ($\lesssim$30\%) saturated in the F606W and F814W
filters.  The short exposures were used for photometry of Proxima itself, and
the long exposures were used for photometry of the background stars. Even though
we have used a large aperture to take the charge bleeding into account in our
photometry in the saturated images (Gilliland 1994) in the F814W and F606W
filters, the photometric uncertainties in these filters are expected to be
higher. The photometric magnitudes for Proxima (on the Vega-mag system) and the
source stars along with the estimated errors are listed in Table~1. We also
examined the 2MASS source catalog for the infrared magnitudes of the sources.
The 2MASS catalog shows a single source at this position, with magnitudes of
$J=16.089$, $H=15.436$, and $K_s=15.764$, and uncertainties of about
$\pm$0.07~mag. These magnitudes most likely represent the combined light of both
background sources. 

Figure~2 shows our long-exposure \HST\/ image taken in the F555W ($V$) filter.
Also shown is the future path of Proxima (green line), calculated using the
\Hipp\/ proper motion and parallax (ESA 1997), and taking into account the
stellar positions as observed by \HST\/ on 2012 October~1. The uncertainty in
Proxima's path is less than the width of the line during the time interval shown
here. As the figure shows, Proxima will actually pass {\it between\/} the two
background stars labeled as ``Source 1" and ``Source 2," affording two
independent opportunities to measure the relativistic light deflection and
search for effects of planetary companions. The closest passages will occur in
2014 October, with an impact parameter of $1\secpoint6 \pm 0\secpoint1$, and in
2016 February, with an impact parameter of $0\secpoint5 \pm 0\secpoint1$. This
prediction is confirmed through a second set of \HST\/ observations that we
obtained in 2013 March. Details of the predicted close encounters are given in
Table~1.  (We have an approved program to obtain further \HST\/ observations at
10 different epochs during 2014 to 2017. The predictions given in Table~1 will
improve as we obtain further \HST\/ observations, and the improved predictions
will be made available at this website: {\tt http://www.stsci.edu/$^\sim$ksahu}.
The exact time of the future \HST\/ observations will be adjusted based on
results obtained from previous observations, in order to maximize the scientific
return.)

We note from Table~1 that the colors of both sources are similar to or bluer than
Proxima, but they are at least 8.5~mag fainter than Proxima. That would imply that
the distances to these source stars are $>$60 pc, much larger than the distance to
Proxima. In our current analysis of predicting the closest approaches, we have
ignored the parallax and proper motion of the source stars, which are expected to
be small. This can be rectified if our future \HST\/ observations show any parallax
or proper motion.

\section{\bf Measuring the Mass
     of Proxima Centauri through Astrometric Microlensing}

When a background star (the source) passes within or close to the Einstein ring of
a foreground object (the lens), it splits into two images and the combined
brightness exceeds that of the source star as depicted in Figure~3. 
As explained in more detail later, the sources will pass 
several Einstein ring radii from Proxima itself,
but could pass very close to or within the Einstein ring of a planet. So we provide  a
general formalism here, which is valid for small as well as large impact parameters.
(For more details, see Schneider, Ehlers, \& Falco 1992; Paczy\'nski 1996; and
Dominik \& Sahu 2000).

The angular radius of the Einstein ring of a point lens can be written as
\begin{equation} 
\theta_\mathrm{E} =  \bigg( { 4\,GM \over {{c^2 d_\pi }} }\bigg) ^{1/2}  , 
\end{equation} 
where $M$ is the lens mass and  $d_\pi$ is the parallax distance of the lens 
as measured with respect to the more distant source, 
given by 
\begin{equation}
{1\over {d_\pi}} =  {1\over D_L} - {1\over D_S} \, ,
\end{equation}
$D_L$ and $D_S$ being the distances from
the Earth to the lens and the source, respectively.   

If the angular separation between the lens and the (undeflected) source 
is $\Delta\theta$, then the two images of the source will be located at
separations of 
\begin{equation}
{\theta}_{\pm} = 0.5 \, \bigg[u \pm \sqrt{u^2 + 4}\bigg] \theta_\mathrm{E}
\end{equation}
relative to the position of the lens, and
\begin{equation}
u \equiv \Delta\theta / 
\theta_\mathrm{E}. 
\end{equation}
Note that the source, the lens, and the two images of the source all lie on a
straight line; the positive sign 
implies the same side as
the source from the lens, and the negative sign implies the opposite side. The angular
distances of the two images from the original position of the source are given
by $({\theta}_{\pm} - \Delta \theta)$. 

Figure 3 schematically shows images as the source (shown by the red dots) passes
though the Einstein ring (blue circle) of the lens. The two images are shown in
black, one outside the Einstein ring (major image),  and the other inside (minor
image).  

The amplifications of the two images of the source are given by
\begin{equation} 
A_{\pm} = 0.5 \, \Bigg[ {{u^2 +2} \over{u \sqrt{4+u^2}}} \pm 1 \Bigg] \, . 
\end{equation}

The combined amplification can be written as
\begin{equation} 
A =  A_+ + A_- = {{u^2 +2} \over{u \sqrt{4+u^2}}}  \, .
\end{equation}
For $u \ll 1$, the combined amplification can be approximated as
\begin{equation} 
A \simeq {1\over u} \, ,
\end{equation}
and for  $u \gg1$
\begin{equation} 
A \simeq {1 + {2\over u^4}} \, .
\end{equation}

The intensity-weighted centroid of the two images is given by
\begin{equation}
{\theta} = {{A_+ {\theta}_+ + A_- {\theta}_-} \over {A}} \, .
\end{equation}
If the two images are unresolved, then the centroid shift with respect to the original  
position of the source can now be written as
\begin{equation}
\delta \theta = {\theta} - \Delta \theta = {u \over {u^2 + 2}} \ \theta_\mathrm{E} \, .
\end{equation}

The locations of the two images, and of their centroid, as functions of $u$ are
shown in the top panel of Figure~4, and the corresponding magnifications are
shown in the bottom panel. 
Note that $\delta \theta$ increases with $u$ when $u < \sqrt{2}$, after which 
it decreases with $u$.
As the figure shows, when $u \gg 1$ (as in the case
of Proxima), $A_- \simeq 0$ and $A \simeq A_+$, so that the position of the
major image can be taken as the centroid of the combined image, and its 
amplification as the amplification of the combined image. 

For $u \gg 1$, the centroid shift can now be
written as
\begin{equation}
\delta\theta \simeq {\theta_\mathrm{E} \over {u}} = {{\theta_\mathrm{E}^2} \over
  {\Delta\theta}} \, .  
\end{equation}

Adopting a distance of 1.3~pc  and a mass of $0.12\,M_\odot$ for Proxima,
and assuming $D_S \gg D_L$ in Eqn.~1, we find that its Einstein
ring radius is $\theta_E \simeq 28$~mas.Thus a
background source lying at an angular distance $\Delta\theta$ from Proxima will
have its apparent position shifted by  
\begin{equation}
{\rm \delta \theta \approx { {(28 ~ mas )^2 \over{\Delta\theta}} }   } \, .
\end{equation}

Using Eqns.~1 and 11, we can express the mass of the lens as
\begin{equation}
M = {{\theta_\mathrm{E}^2 c^2 d_\pi} \over
  {4 G}} \simeq
{{ \delta\theta \, c^2 d_\pi \Delta\theta} \over 4G} \, .
\end{equation}

The angular separation $\Delta\theta$, and the centroid shift  $\delta\theta$
can be determined from the observations taken before and during the event. To
measure the mass of the lens using Eqn.~13, the remaining required parameter is
the parallax distance $d_\pi$, depending only on the {\em difference\/} between the
parallaxes of the lens and the source. The parallax of the lens is already
known, and our planned \HST\/ observations will directly constrain the source
parallaxes (expected to be very small as noted above).

For the upcoming close encounters of Proxima Centauri, the sources lie on either
side, with impact parameters of 0.5 and 1.6 arcsec, which correspond to  $u
\simeq 18$ and 57, respectively. For such large $u$, the major image  will lie
within $\sim$1.5~mas of the undeflected source location. The brightness of the
minor image, which will lie within $\sim$1~mas of Proxima itself, will be $<$$ 2
\times 10^{-5}$ that of the primary image, and thus can be ignored.     

In Figures~5 and 6 we plot the centroid shifts of Sources~1 and~2 as functions
of time, under the assumption that Proxima has a mass of $0.12\,M_\odot$\null.
The maximum centroid shift for Source~1 is $\sim$0.5~mas at closest approach,
and for Source~2 it is $\sim$1.5~mas.  As seen from Eqn.~12, the centroid shift
scales inversely with the angular separation between Proxima and the source
star,  so the deflections are already underway as we write!  Assuming that the
centroid shifts of the two sources can be measured with an accuracy of 200
mas per epoch, Proxima's mass can be measured with an accuracy of
$\sim$5\%.

\section{Astrometric Detection of Planets around Proxima}

If Proxima has planetary companions, it may be possible to detect them---and
measure their masses---through their extra astrometric shifts of one or
both of the background stars (Safizadeh, Dalal, \& Griest 1999;
Han \& Lee 2002).  Using Eqn.~1, we can write the size of the
angular Einstein ring due to a planet around Proxima as
\begin{equation}
(\theta_\mathrm{E})_{\rm planet} \simeq 8 \ \Bigg({M_{\rm planet} \over 
  M_{\rm Jup}}\Bigg)^{1/2} \, {\rm mas} \, . 
\end{equation}
The maximum possible extra deflection due to a Jupiter-mass planet is $(\sqrt
2/4) \, \theta_\mathrm{E} \simeq 2.8$~mas, which would occur when the angular separation
between the background source's deflected image and the planet is $\sqrt 2 \
\theta_\mathrm{E} \simeq 11$~mas. 

Figure~6 shows an example of how the gravitational deflection of Source~2 might be
modified by the presence of a planetary companion of Proxima under favorable
circumstances. We have assumed a Jupiter-mass companion in a face-on circular
orbit with a separation of 0.8~AU, which has a closest approach of 14~mas from
the background star. The dashed lines in both panels show the deflections when
this hypothetical planet is included in the calculations. The effect of the
motion of the planet in its 751-day orbit is included. As the figure
illustrates, such a planet would produce a relatively large distortion of the
deflection curve by about 0.5~mas, but lasting only about 3~days. 

Let us now calculate the sizes of the regions around Proxima in which planets
can be detected through their astrometric signals.  If the minimum astrometric
deflection that can be measured is $\delta_{\rm min}$, then the corresponding
maximum angular distance between the source and the planet can be written, using
Eqn.~9, as
\begin{equation}
\phi_{\rm max} = 
  {{(\theta_\mathrm{E})_{\rm planet}^2} \over {\delta_{\rm min}}} \simeq 
  {{(8\,{\rm mas})^2} \over  {\delta_{\rm min}}} \,
  \Bigg(  {{M_{\rm planet}} \over {M_{\rm Jup}}}  \Bigg) \, . 
\end{equation}
This equation implies, for example, that if $\delta_{\rm min} = 0.03$~mas 
(detectable with \HST\/ spatial scanning as discussed above),  the deflection
due to a Jupiter-mass planet would be detectable if the planet passes within
$2''$ of either background source. 

Figure~7 illustrates the spatial regions around Proxima in which planets of
various masses will be detectable during the upcoming approaches of the two
background stars. Here we show the trajectories of the two stars in a
Proxima-centered reference frame. Dotted curves refer to Source~1, and solid
curves to Source~2. The sets of blue curves surrounding both trajectories, with
separations of $\sim$$2''$ from the stars, enclose the regions in which a
Jupiter-mass planet is detectable with a deflection exceeding 0.03~mas. The
red curves surrounding the two trajectories enclose the regions in which the
deflection due to a $10\,M_{\rm Earth}$ planet is detectable.

Even though astrometric microlensing provides an exciting possibility of
detecting planets around Proxima, the probability of such a detection is small
for two reasons. First, the microlensing signal due to a planet lasts only about
2~days. Second, there is a zone close to Proxima where it becomes difficult to
distinguish the signal due to a planet from that due to Proxima itself.
Detection of the planetary signal is easier when the planet passes closer to the
source star, and is most favorable when the planet passes through a point where
its separation from the source is much less than the planet's separation from
Proxima.

\section{Photometric Microlensing by Planets around Proxima}

If Proxima has a planet which passes within a few mas of either source,
photometric signatures of microlensing due to the planet may be detectable.  As
noted earlier, the Einstein ring of a Jupiter-mass planet has a radius of about
8~mas. For a photometric accuracy of 1\%, the planetary signature will be
detected for an amplification of $A=1.01$, which is achieved if the impact
parameter $u\le3.5$ (Eqn.~4). This requires the planet to lie within an angular
distance $\zeta$ of the source, where
\begin{equation}
\zeta = 3.5 \, (\theta_\mathrm{E})_{\rm planet} \simeq 28 \,  
 {\Bigg( {M_{\rm planet}\over{M_{\rm Jup}}} \Bigg)^{0.5}} \, \rm\, mas.
\end{equation}
Thus any Jovian-mass planet that passes within 28~mas of either background star
(corresponding to a linear separation of $\le$0.036~AU) can be detected through
precision photometry.  

The Einstein-ring crossing time is given by $T_\mathrm{E} = \theta_\mathrm{E}/A$, where $A$ is
the proper motion of the planet relative to the background star. Assuming the 
proper motion of Proxima for the planet, the crossing time is
\begin{equation}
(T_E)_{\rm planet} \simeq 2.7 \,
  {\Bigg( {M_{\rm planet}\over{M_{\rm Jup}}} \Bigg)^{0.5}} \, \rm days. 
\end{equation}
The duration of the photometric event can be written as 
\begin{equation}
T_\mathrm{D} = 2 \, T_\mathrm{E} \, (b^2 - u^2)^{0.5} \, ,
\end{equation}
where $b$ is the source-lens separation, in units of $\theta_\mathrm{E}$, at which  the
amplification is detectable. An amplification of $A=1.01$ corresponds to $b =
3.5$. Thus $T_\mathrm{D}$ can be up to 7 times $T_\mathrm{E}$, or, for a Jupiter-mass planet,
$\sim$19~days. 

In Figure~7 the green lines surrounding each background source trajectory
enclose the regions within which a 1\% photometric amplification by a Jovian
planet would be detectable.  These zones are very small, so the probability
of detecting photometric microlensing by a planet is extremely small.

\section{Observational Considerations}

Successful observations of these events will be challenging, because the
background stars are $\ge$8.5~mag fainter than Proxima and will require accurate
astrometry and photometry at separations from Proxima ranging from $0\farcs5$ to
$\sim$$2''$. Here we discuss the feasibility of the observations from space and
from the ground.

\subsection{Space-based Astrometry and Photometry}

For the upcoming events, the minimum values of $\Delta\theta$ will be
$1\secpoint6$ and $0\secpoint5$, corresponding to maximum deflections of 0.5 and
1.5~mas. Such displacements are routinely measured using cameras onboard \HST\/;
in fact, astrometric accuracies of 0.2~mas have been achieved in single
measurements through direct imaging with the Advanced Camera for Surveys (ACS)
and WFC3 (e.g., Anderson et al.\ 2006; Bellini et al.\ 2011), and
spatial-scanning modes have been developed recently with WFC3 that can achieve
astrometric accuracies of about 0.03~mas (A.~Riess, private communication).

The fact that these observations can be obtained easily with \HST\/ even in the
presence  of Proxima is illustrated in Figure~8. In order to achieve astrometric
accuracy of 0.2~mas per observation, we need a S/N of about 300 for each
background star, which can be achieved in an exposure time (WFC3, F555W filter)
of 70~s. The left panel shows a 200-s exposure on Proxima taken with WFC3 in
F555W filter, which is thrice the required exposure time. The white circle has a
radius of 12.5 pixels ($0\farcs5$).  The right-hand panel shows the radial
profile of this star, greatly magnified to show the signal in the wings of the
stellar image. The counts drop below 5000~electrons at a radial distance of
12.5~pixels, at which point the background becomes negligible
compared to a star with S/N = 300. Since the source star will actually be at a
separation of $0\secpoint5$ even at closest approach, saturation effects are
thus unimportant provided the diffraction spikes and charge bleeding are
avoided.

These events will conveniently occur during the operational life time of {\it
Gaia}, ESA's astrometry mission, recently successfully launched.  {\it Gaia\/}
can observe down to $V\simeq 20$, and its astrometric accuracy is expected to be
0.1 to 0.3 mas depending on the source magnitude (Gare et al.\ 2010; Prusti
2012; Liu et al.\ 2012). Assuming {\it Gaia\/} can observe the fainter
background star in the presence of  the much brighter Proxima, the mass of
Proxima should be measured very well.

\subsection{Ground-based Astrometry and Photometry}

Photometric observations required for this project are possible with several
ground-based instruments. One example is the upcoming SPHERE 
(Spectro-Polarimetric High-contrast Exoplanet REsearch) of ESO VLT, which is
specially designed to image faint sources  in the presence of sources brighter
by as much as 12.5 magnitudes with separations as small as $0\secpoint1$ (Beuzit
et al.\ 2008).  The Gemini Planet Imager may also be a suitable instrument where
a contrast of  $>$$10^6$ at $0\secpoint 4$, and closed-loop performance  down to
$I=8$ mag, have been achieved  (Macintosh et al.\ 2012). The ``lucky imaging
system" at the Danish 1.54~m telescope at the ESO La Silla Observatory  may also
be a capable instrument, which uses a EMCCD (Electron-multiplying Charge Coupled
Device) that delivers fast readout times and negligible readout noise, making it
an ideal detector where a large number of fast frames can be used in ``shift and
add" mode to improve resolution  and avoid saturation (Harps\o e et al.\ 2012).

Astrometric observations from the ground are, however, very challenging.
Astrometric accuracies of 0.1~mas have been achieved for stars brighter than
$K=16$ with NaCo (Fritz et al.\ 2009; Lenzen et al.\ 2003; Rousset et al.\
2003), which is a Near-IR Imager and spectrograph with an adaptive optics system
at ESO VLT.  Astrometric measurements with an accuracy of 0.1 mas for sources
brighter than $K=16$ have also been achieved with Keck (Clarkson et al.\ 2012),
although Proxima's declination of $-62$ degrees makes it unobservable.
However, apart from the fact that both our sources are  most likely fainter than
$K=16$, they will be accompanied by a very close and much brighter star. In
principle, the brighter star can help by serving as a good guide/reference star,
but in practice, the ``halo" of the bright star  is likely  to overlap with the
position of the target star (Fritz et al.\ 2009), posing challenges
by causing systematics which need to be characterized extremely well.  

\section{Summary}

We have shown that the nearest star, Proxima Centauri, will pass close to two
background stars in 2014 and 2016, with impact parameters of about $1\farcs6$
and $0\farcs5$. Because Proxima is so nearby, its angular Einstein ring radius
is large ($\sim$28~mas) and will lead to detectable relativistic light deflections
of the images of the background stars even at those angular separations. 
Measurement of the astrometric shifts offers a unique opportunity for the 
determination of the mass of Proxima with an accuracy of $\sim$5\%. 

Although the background stars are $\ge$8.5~mag fainter than Proxima, the large
contrast is mitigated by the relatively large separations at which the
gravitational deflection is still detectable, and well within the capabilities
of the {\it Hubble Space Telescope\/}.

The upcoming events also offer the opportunity to detect and determine the
masses of planetary companions, either through additional astrometric shifts, or
in rare circumstances through a photometric microlensing event, leading to a
brightening of the source star. These events would have durations of a few hours
to several days. 

\acknowledgments

We thank R\'emi Soummer, Will Clarkson, and Rosanne Di Stefano for useful
discussions.  Partial support for this research was provided by NASA through
grant GO-12985 from the Space Telescope Science Institute, which is operated by
the Association of Universities for Research in Astronomy, Inc., under NASA
contract NAS5-26555. KS acknowledges support from the European Southern
Observatory, the Institute for Theory and Computation at the Harvard-Smithsonian
Center for Astrophysics, and the Institute for Advanced Study at Princeton for
sabbatical visits, during which parts of this work were carried out.  MD is
thankful to Qatar National Research Fund (QNRF), member of Qatar Foundation, for
support by grant NPRP 09-476-1-078.



{\it Facilities:}  \facility{HST (WFC3)}.




\begin{table}
\begin{center}

\caption{Details of the source stars.\label{tbl-1}}
\vskip 0.2cm
\begin{tabular}{lccc}
\tableline
 Parameter & Source 1 & Source 2 & Proxima\tablenotemark{a}\\
\tableline
RA (J2000)&          14:29:34.693&  14:29:34.268 \\
Dec (J2000) &       $-62$:40:33.46&  $-62$:40:34.91 \\
F475W (``$B$")& 21.26 & 20.55 & 12.03 \\
F555W  (``$V$")&         20.36 &         19.89 & 11.33 \\
F606W (``wide-$V$")& 19.61 & 19.29 & (10.44$^a$) \\
F814W (``$I$")& 17.78 & 17.93 & (7.25$^a$) \\
Date of closest approach & $2014.80 \pm 0.03$ &  $2016.16 \pm 0.03$ \\
& (2014 Oct 20  $\pm$ 10 days) & (2016 Feb 26  $\pm$ 10 days) \\
Impact Parameter   &   $1\secpoint6 \pm 0\secpoint1$ & $0\secpoint5 \pm 0\secpoint1$  \\
\tableline
\end{tabular}
\tablenotetext{a}{The magnitudes of Proxima in the F606W and F814W filters
are determined from saturated images which are  
uncertain by $\pm$0.5 mag. All other magnitudes have uncertainty of 
$\pm$0.05 mag.}

\end{center}
\end{table}

\begin{figure}
\begin{center}
\includegraphics[height=3.75in]{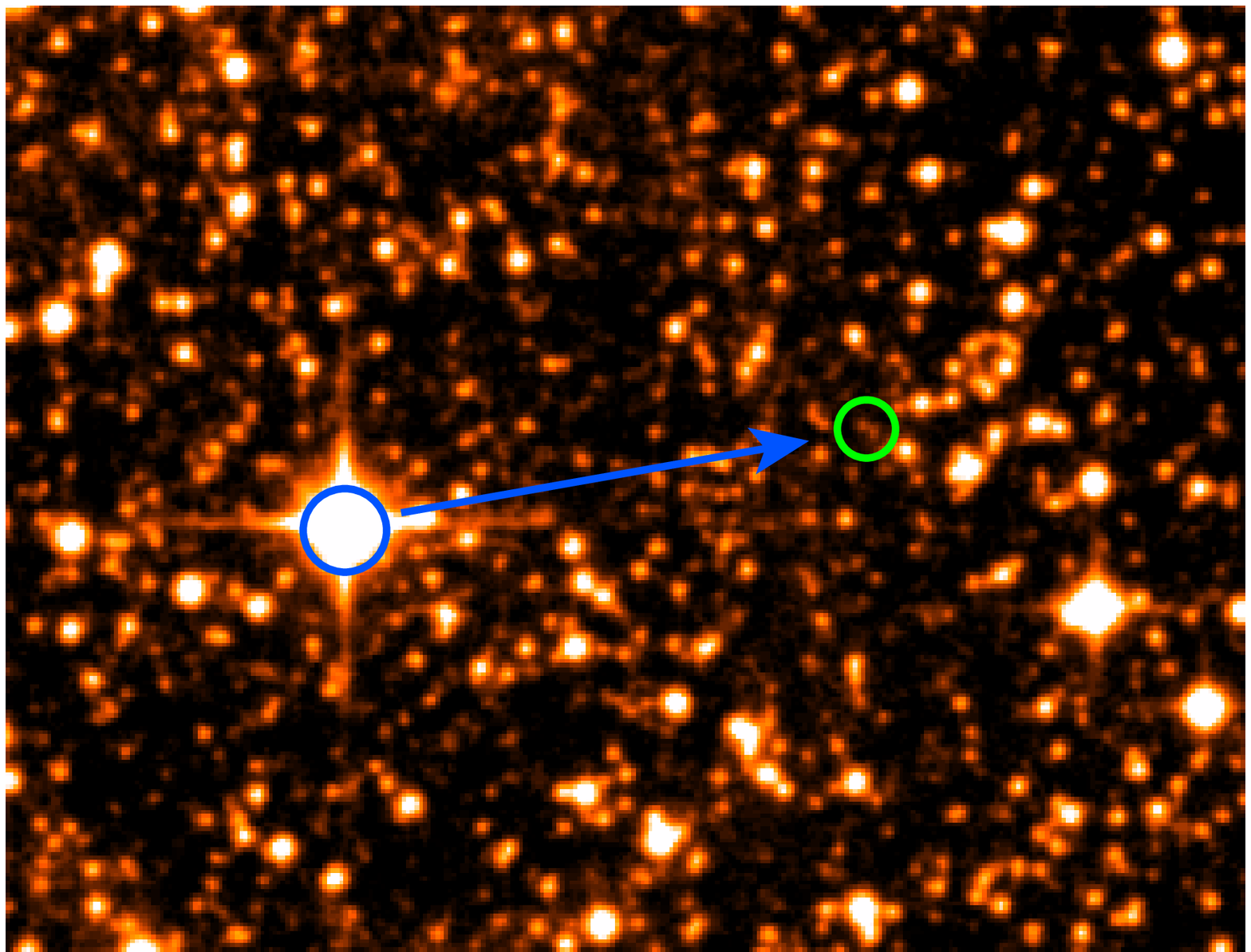}
\vskip0.1in 
\includegraphics[height=3.75in]{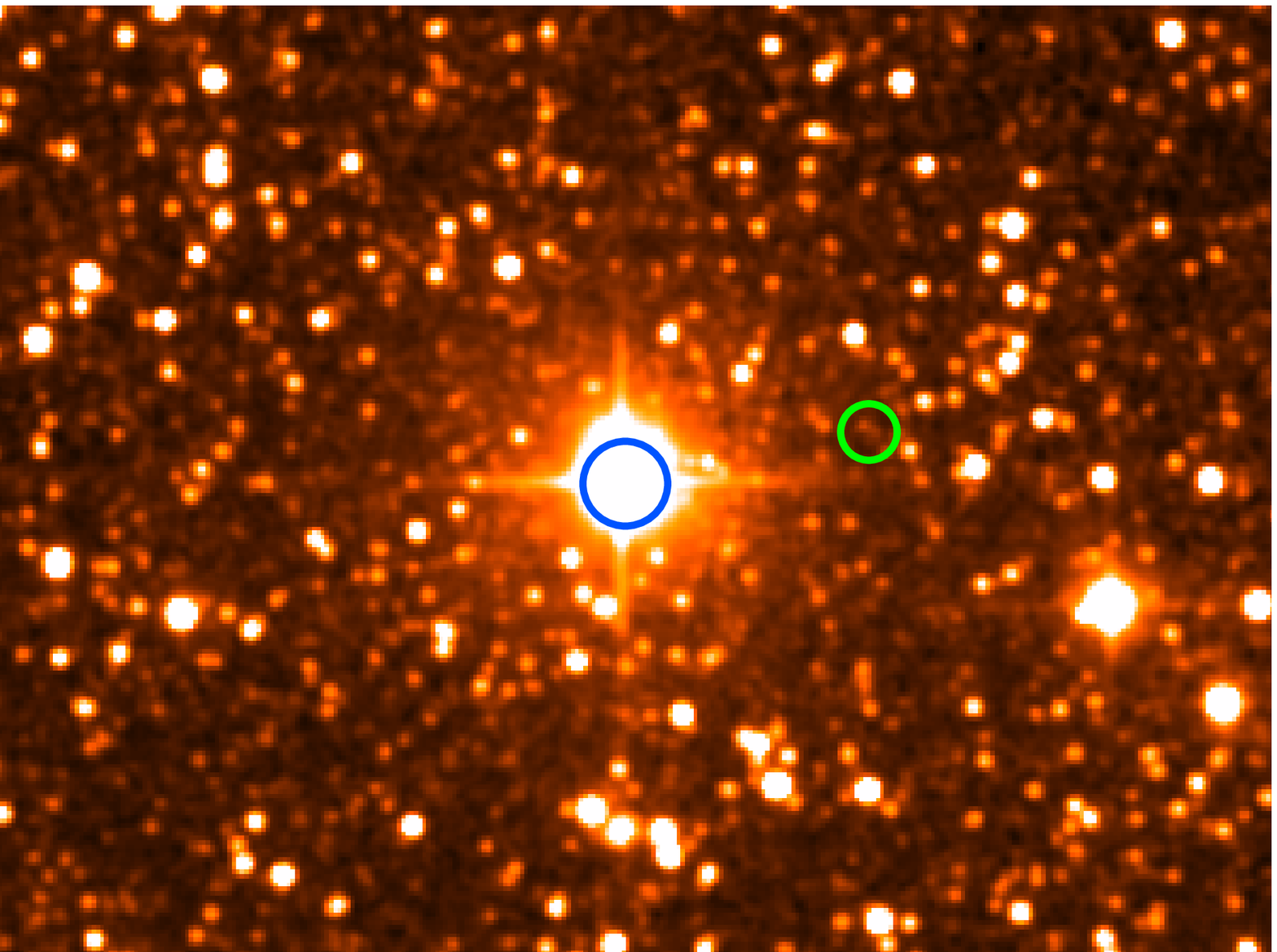}
\end{center}
 \caption{{\bf (Top):} Proxima Centauri field in 1997, taken from the near-IR
Digitized Sky Survey.  The image size is $300'' \times 220''$ with north at the
top and east on the left.  Proxima is encircled in blue, and is moving to the
west-northwest. {\bf (Bottom):} Proxima field in 2000, taken in the 2MASS
survey. In both images the pair of faint background stars that will be
microlensed in 2014 and 2016 is within the green circles on the
right.
}  
\label{deltatraj} 
\end{figure}

\begin{figure}
\epsscale{.80}
\plotone{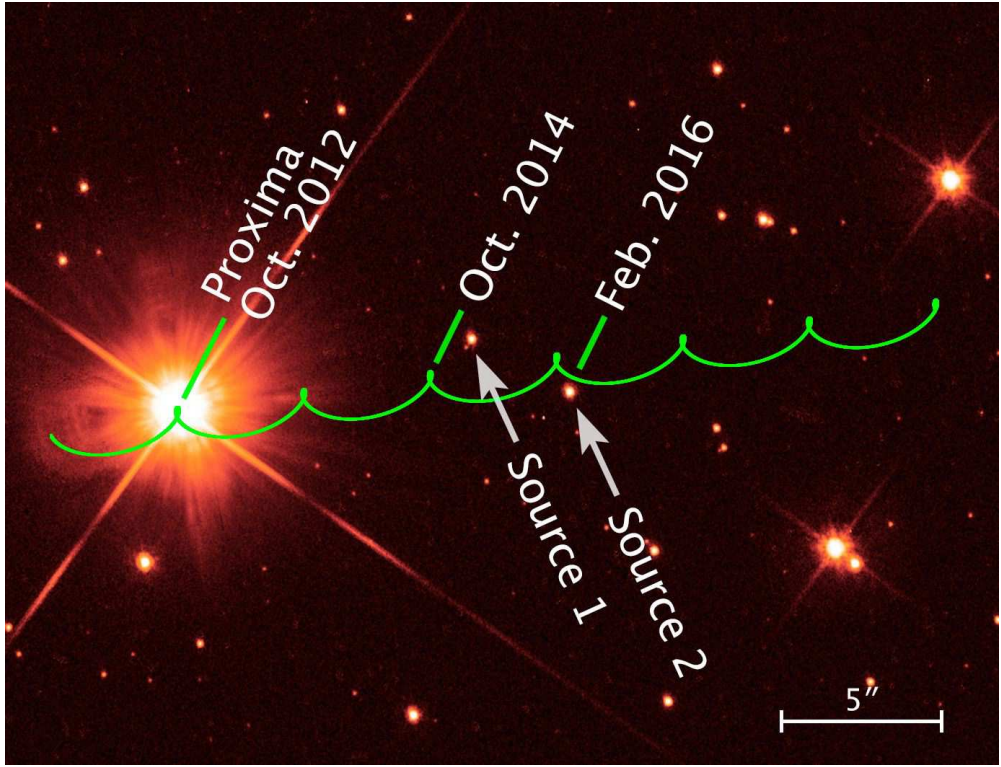}
\caption{Proxima Centauri field as observed with \HST/WFC3 on 2012 October~1 in the
F555W ($V$) filter, with north up and east to the left.   The two faint background
sources are labelled ``Source 1" and ``Source 2."  The future path of Proxima is
shown in green, taking into account proper motion and parallax.  Considering
the facts that (i)~the proper motion of Proxima has an accuracy of $\sim 2$
mas/yr,  (ii)~the direction of proper motion has an uncertainty of $0\fdg036$, and
(iii)~Proxima was observed by \HST\/ in 2012 October, the uncertainty in the proper
motion is less than the width of the trajectory during the time interval shown in
the figure. The locations and dates of closest approach of Proxima to the two
background  sources are indicated. The closest approaches to the two background
stars will occur in 2014 October and 2016 February, at separations of  $1\secpoint6
\pm 0\secpoint1$ and $0\secpoint5 \pm 0\secpoint1$, respectively (see Table 1).
Note that there is an additional  faint source about $0\secpoint3$ southeast of
Source 1, which is  about 2.7~mag fainter than Source 1 in the F555W filter.}
\end{figure}

\begin{figure}
\epsscale{.80}
\plotone{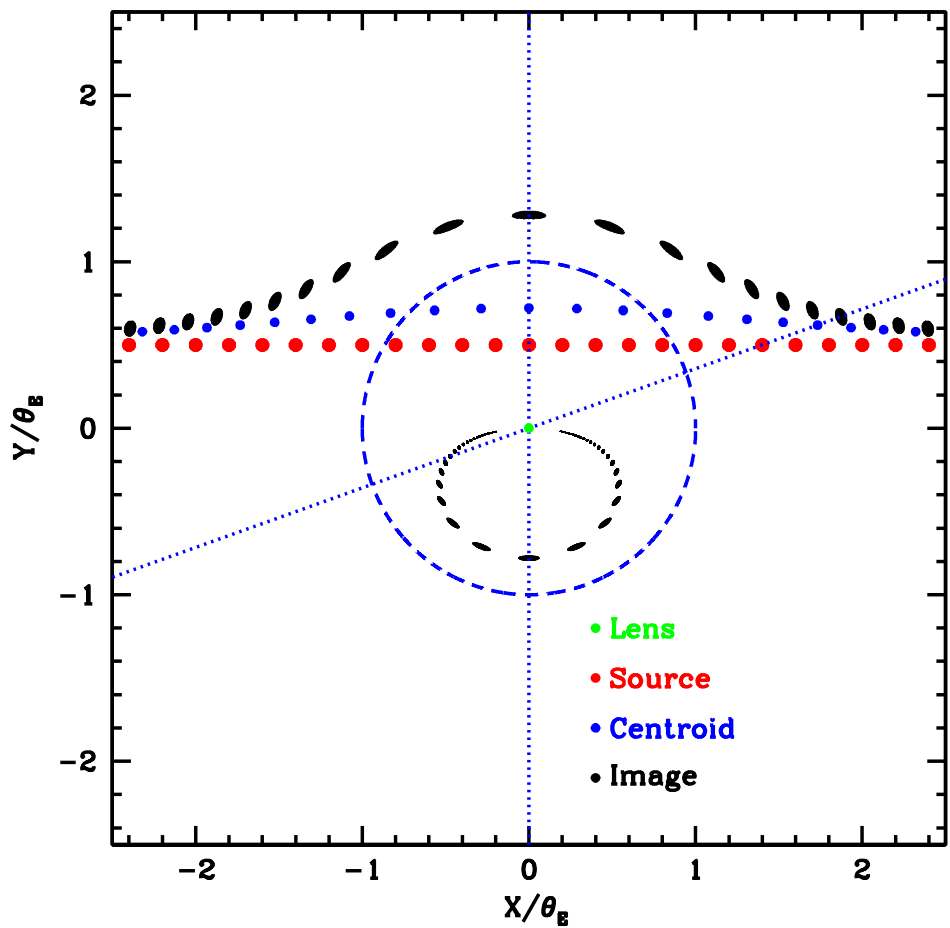}
\caption{This figure shows how the apparent positions and the sizes of the images
change at various stages of a microlensing event. In this geometry the position of
the lens, indicated by a green dot, is fixed, and the red dots show the actual
positions of the source. The black ellipsoidal points show the images of the source as it
passes close to the lens in the plane of the sky. The dashed circle is the
Einstein ring of the lens.  At any instant, the source, the lens and the two
images lie on a straight line as shown by the dotted blue lines.  The centroid of
the  images are indicated by the blue dots, which are clearly shifted with respect
to the source positions due to microlensing. The maximum centroid shift occurs
when the lens-source separation is $\sqrt 2 \theta_E$; one such case
is shown by a blue dotted line (cf.\ Paczy\'nski 1996).}
\end{figure}

\begin{figure}
\epsscale{.60}
\plotone{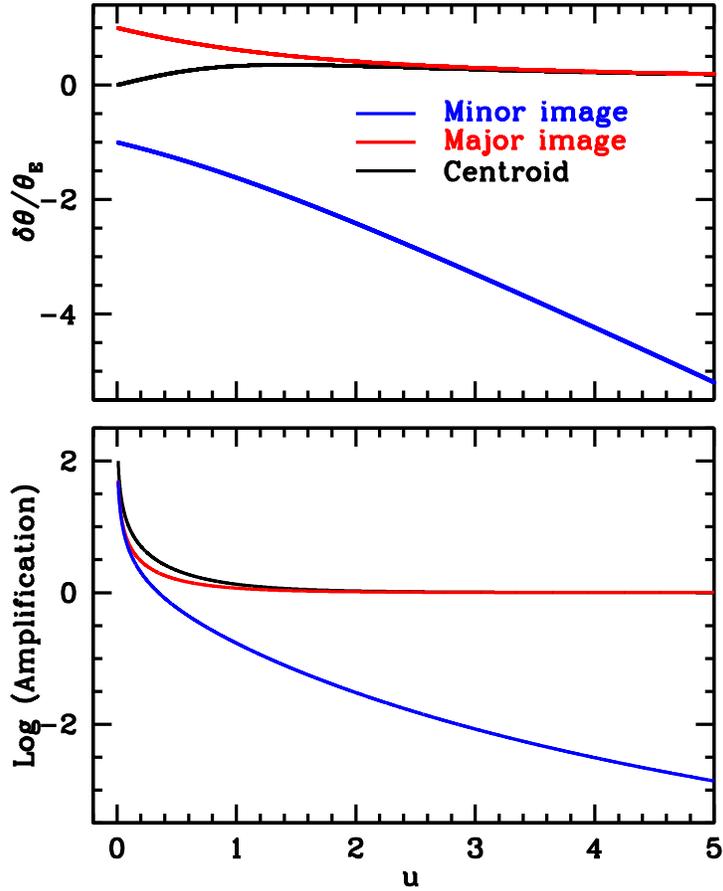}
\caption{The astrometric shift and the amplifications of the images. The upper
panel shows the shifts of the minor image, the major image and the centroid
with respect to the undeflected position of the source,
as a function of $u$. The lower panel shows the amplifications of the same
three components as a function of $u$.
The shift of the minor image monotonically increases with $u$, but its
brightness drastically reduces for larger  $u$ as seen in the lower panel. 
As a result, the contribution of the minor image is negligible at high $u$, 
and the astrometric shift and the amplification are
close to that of the major image.   
}
\end{figure}

\begin{figure}
\epsscale{1.20}
\plottwo{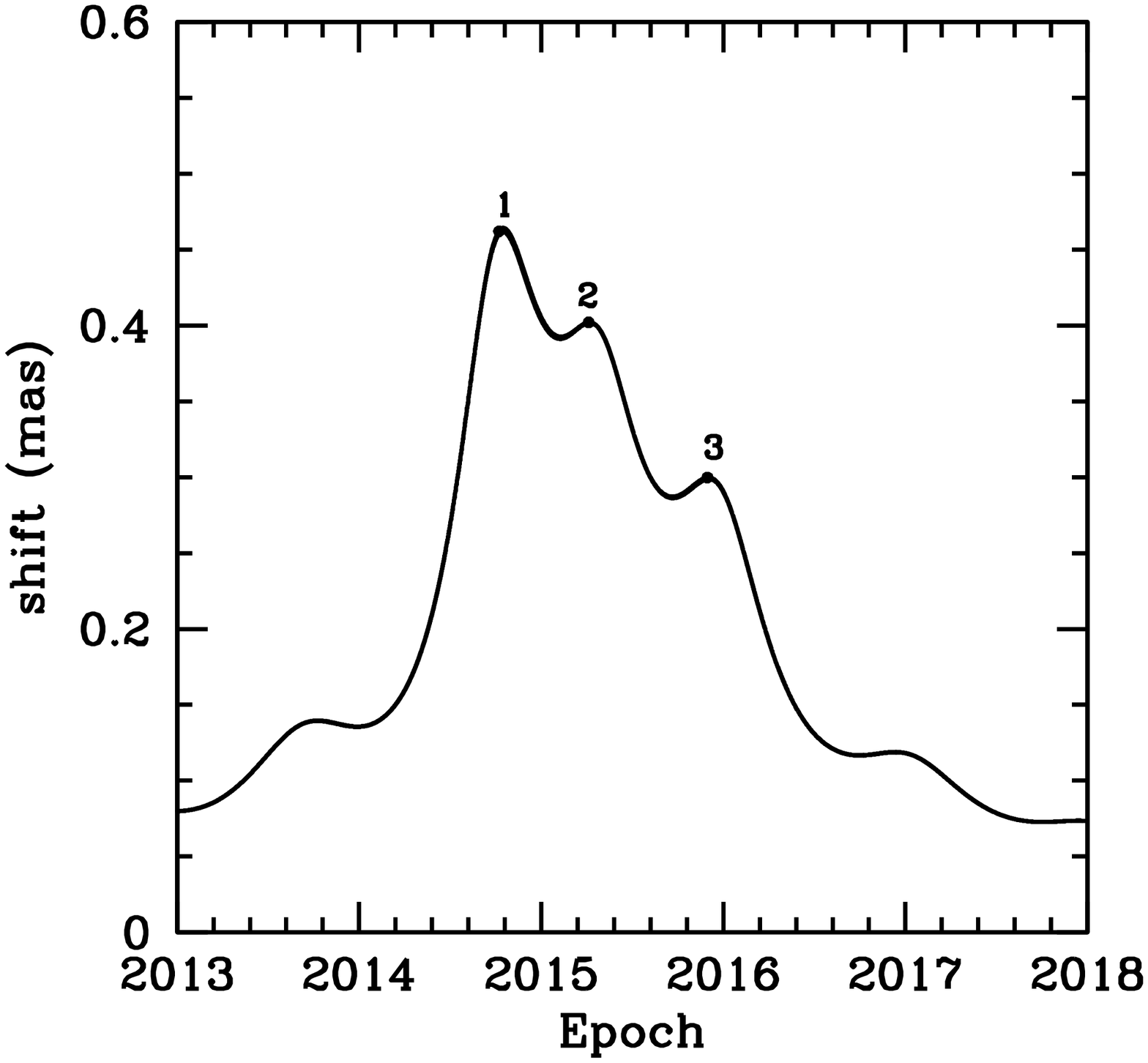}{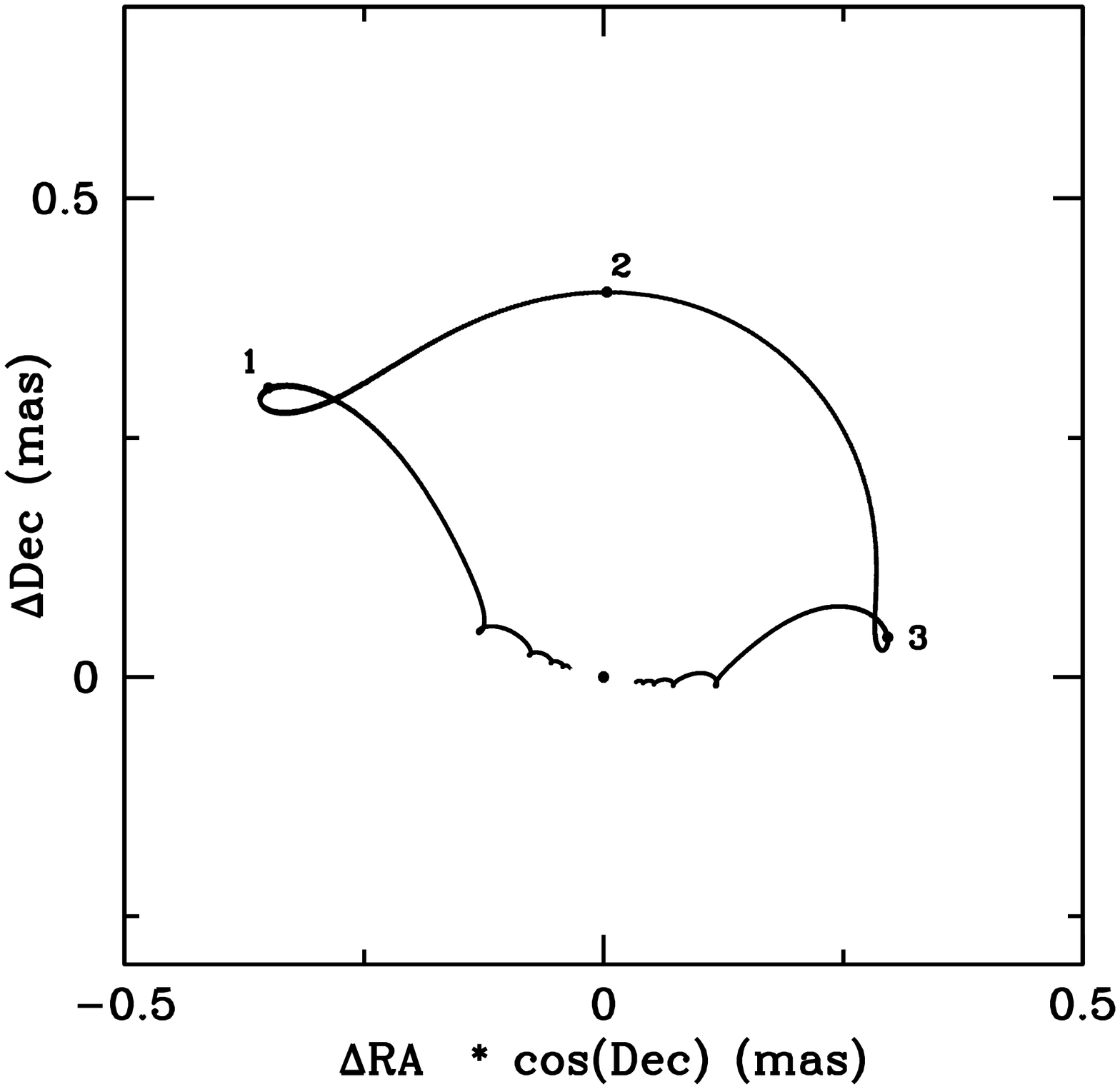}
\caption{Expected astrometric shifts of Source~1 (the fainter of the two
background stars). The left panel shows the total astrometric shift as a
function of time, and the right panel shows the two-dimensional motion of the
image of the background star. The epochs of three peak deflections are marked
with numerals 1, 2, and 3 in the left panel, and the corresponding points are
labeled in the right panel.}
\end{figure}

\begin{figure}
\epsscale{1.20}
\plottwo{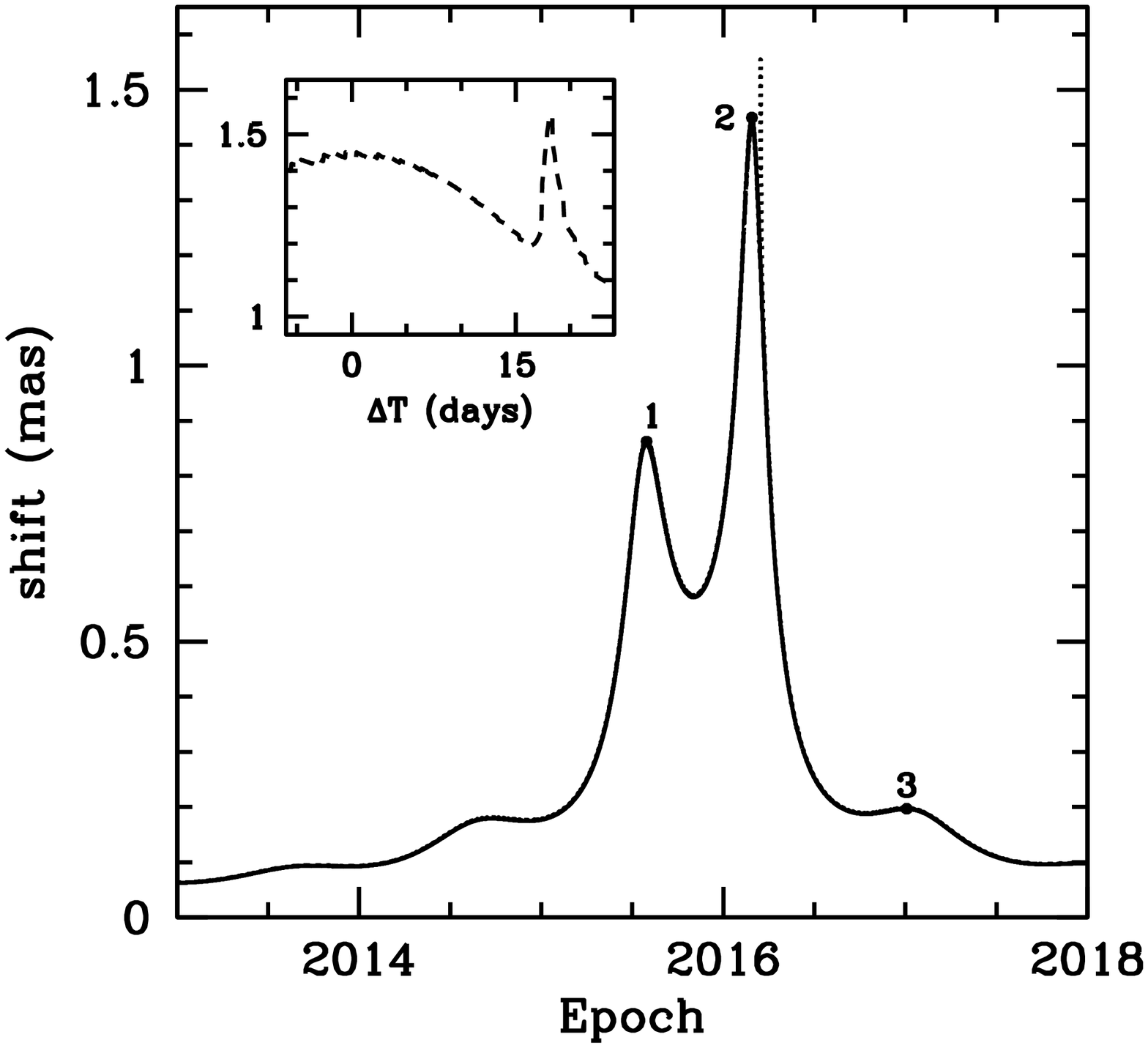}{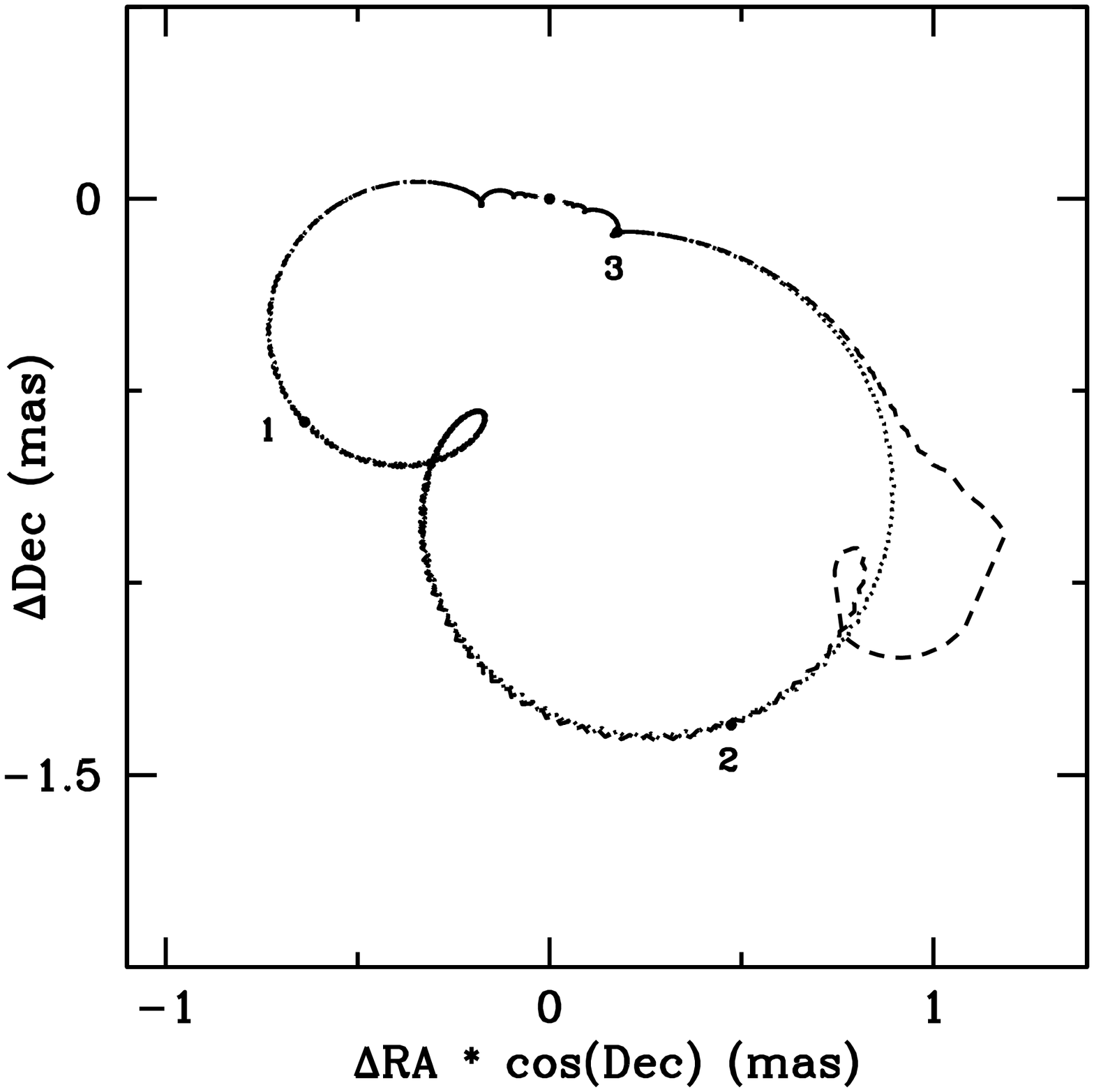}
\caption{Expected astrometric shifts of Source~2 (the brighter of the two
background stars). The solid line in the left panel shows the total astrometric
shift as a function of time, and the right panel shows the two-dimensional
motion of the image of the background star. The epochs of three peak deflections
are marked with numerals 1, 2, and 3 in the left panel, and the corresponding
points are labeled in the right panel. The dashed lines in both panels show the
shifts when we additionally assume Proxima to be accompanied by a Jupiter-mass
planet at an orbital separation 0.8~AU, which passes within 14~mas of the
background star. These calculations take the
orbital motion of the planet into account.  The inset in the left panel shows a
magnified view of the region of the planet's effect on the astrometric shift,
which has a duration of about 3~days.} 
\end{figure}

\begin{figure}
\epsscale{.60}
\plotone{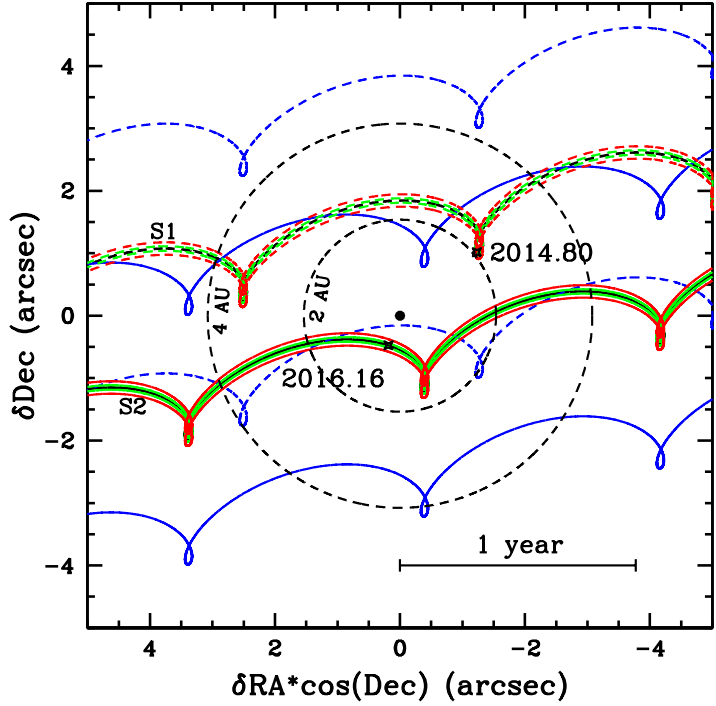}
\caption{Trajectories of the two background stars in a Proxima-centered reference
frame, shown as dashed (Source~1, labelled as ``S1") and  solid (Source~2, labelled
as ``S2") black lines.  Black dot at center denotes Proxima, surrounded by two
dashed black circles of radii 2 and 4~AU\null.  If Proxima has a $1\,M_{\rm Jup}$
planet lying between the pairs of dashed blue (Source~1) or solid blue (Source~2)
lines, the extra astrometric shift caused by the planet will (briefly) exceed
0.03~mas. If Proxima has a $10\,M_{\rm Earth}$ planet lying between the pairs of
red lines, the astrometric signal of this planet will likewise exceed 0.03~mas.  If
Proxima has a $1\,M_{\rm Jup}$ planet within the green lines, its {\it
photometric\/} microlensing  signal will exceed 0.01~mag. The locations and dates
of the closest approaches of the two background stars are marked. As they move to
the east (left) relative to Proxima, the parallactic loops of the background stars
have a one-year timescale, as indicated by the scale bar at the bottom. The x-axis
denotes the absolute value of the shift, i.e $\delta {\rm RA} \times \rm
{cos(Dec)}$.
}
\end{figure}

\begin{figure}
\epsscale{1.2}
\plottwo{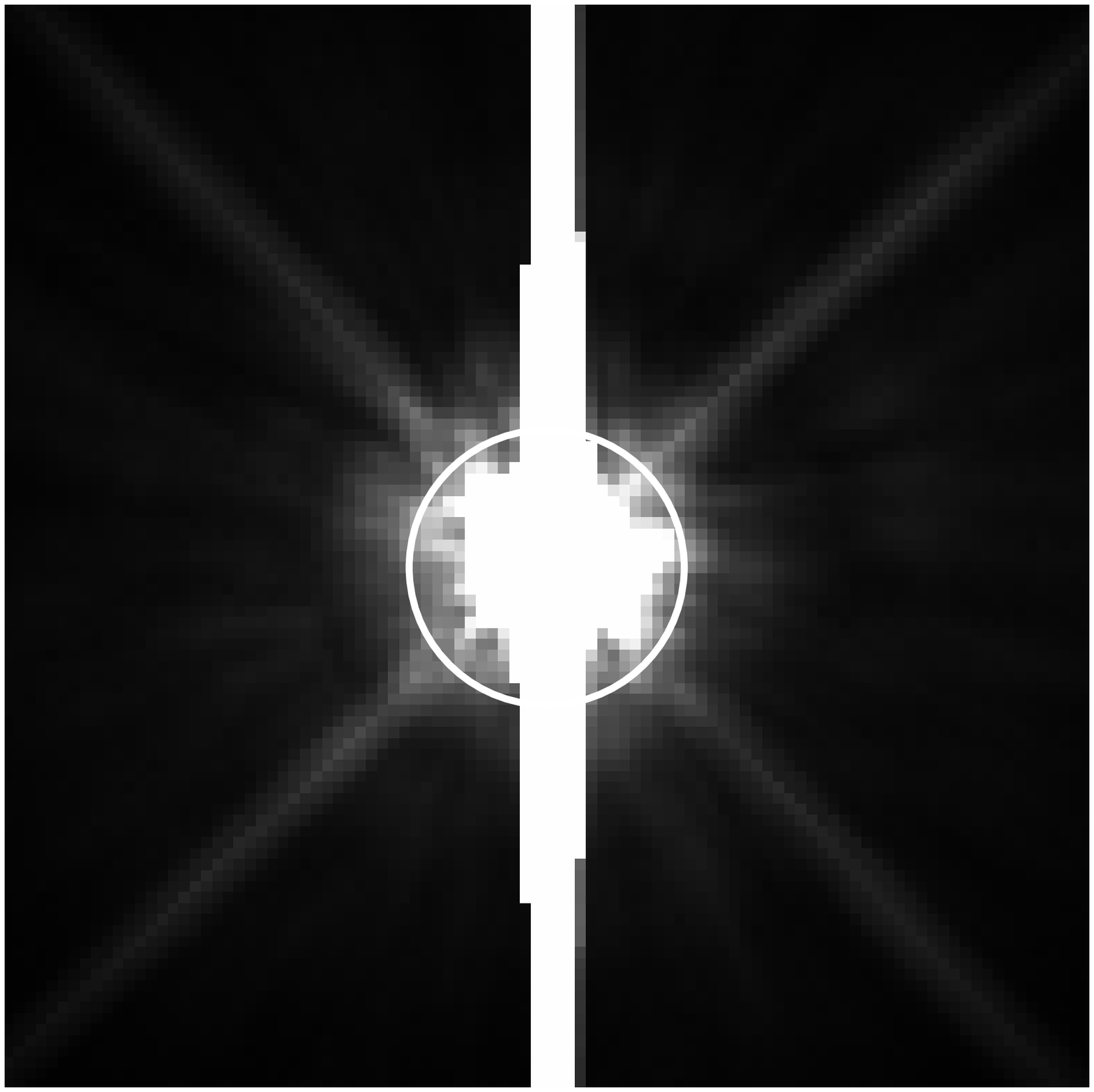}{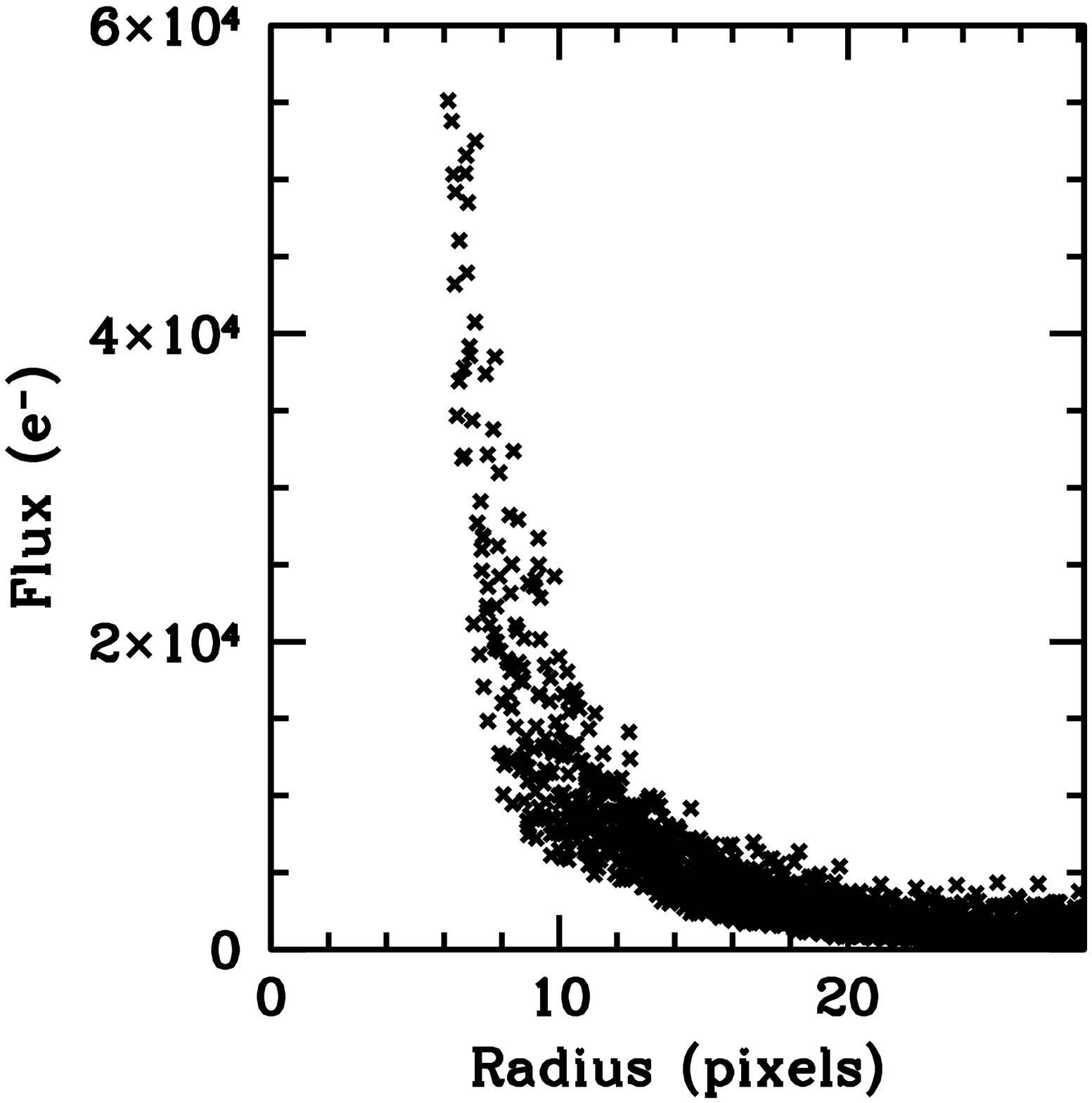}
\caption{
{\bf (Left):} Image of Proxima Centauri taken  with WFC3/\HST\/ on 2012 October
1, in the F555W filter with an exposure time of 200 seconds  This is about 3
times overexposed than what is required  in order to attain a S/N of 300 for
even the fainter background star. The white circle corresponds to a radius of
12.5 pixels corresponding to 0.5 arcsec, which is the closest approach of the
background star. {\bf (Right)}:  Radial profile of Proxima in the 200-sec
image,  greatly magnified to show the signal in the wings of the stellar image.
The counts drop below 5000~electrons at a radial distance of $0\secpoint4$, at
which point the background becomes small compared to a star with S/N = 300.
Since the source star will actually be at a separation of $0\secpoint5$ at
closest approach, it is important to choose an orientation of the camera such
that the sources do not fall either on the direction of charge bleeding,  or on
the diffraction spikes.
  }  
\end{figure}

\end{document}